\documentstyle[12pt]{article}

\def\be{\begin{equation}}       \def\eq{\begin{equation}}
\def\ee{\end{equation}}         \def\eqe{\end{equation}}

\def\bea{\begin{eqnarray}}      \def\eqa{\begin{eqnarray}}
\def\ena{\end{eqnarray}}        \def\eea{\end{eqnarray}}
                                \def\eqae{\end{eqnarray}}

\def\ba{\begin{array}}
\def\ea{\end{array}}
\def\unit{1 \hskip-.3em \raise2pt\hbox{$ \scriptstyle |$ } }

\def\a{\alpha}
\def\b{\beta}
\def\c{\gamma}
\def\d{\delta}
\def\e{\epsilon}           
\def\f{\phi}               
\def\vf{\varphi}  \def\tvf{\tilde{\varphi}}
\def\g{\gamma}

\def\m{\mu}
\def\n{\nu}
  
\def\p{\pi}                
\def\r{\rho}                                     
\def\s{\sigma}                                   

\def\x{\xi}

\def\D{\Delta}
\def\F{\Phi}
\def\G{\Gamma}

\def\P{\Pi}

\def\S{\Sigma}

\def\ca{{\cal A}}

\def\cn{{\cal N}}
\def\co{{\cal O}}

\def\half{{1 \over 2}}

\def\bop#1{\setbox0=\hbox{$#1M$}\mkern1.5mu
        \vbox{\hrule height0pt depth.04\ht0
        \hbox{\vrule width.04\ht0 height.9\ht0 \kern.9\ht0
        \vrule width.04\ht0}\hrule height.04\ht0}\mkern1.5mu}
\def\Box{{\mathpalette\bop{}}}                        
\def\pa{\partial}                              
\def\de{\nabla}                                       

\def\>{\rangle} 

\def\<{\langle} 
\def\Dsl{D \hskip-.6em \raise1pt\hbox{$ / $ } }

\def\leftrightarrowfill{$\mathsurround=0pt \mathord\leftarrow \mkern-6mu
       \cleaders\hbox{$\mkern-2mu \mathord- \mkern-2mu$}\hfill
       \mkern-6mu \mathord\rightarrow$}
\def\dvec#1{\vbox{\ialign{##\crcr
       \leftrightarrowfill\crcr\noalign{\kern-1pt\nointerlineskip}
       $\hfil\displaystyle{#1}\hfil$\crcr}}}          
\def\hook#1{{\vrule height#1pt width0.4pt depth0pt}}
\def\leftrighthookfill#1{$\mathsurround=0pt \mathord\hook#1
       \hrulefill\mathord\hook#1$}
\def\underhook#1{\vtop{\ialign{##\crcr                 
       $\hfil\displaystyle{#1}\hfil$\crcr
       \noalign{\kern-1pt\nointerlineskip\vskip2pt}
       \leftrighthookfill5\crcr}}}
\def\smallunderhook#1{\vtop{\ialign{##\crcr      
       $\hfil\scriptstyle{#1}\hfil$\crcr
       \noalign{\kern-1pt\nointerlineskip\vskip2pt}
       \leftrighthookfill3\crcr}}}

\def\to{\rightarrow}
\def\tf{\tilde{\f}}

\def\pa{\partial}

\newcommand{\sm}[1]{\mbox{\scriptsize #1}}

\def\de{\mbox{d}}

\def\nonu{\nonumber \\{}}
\def\half{{1 \over 2}}
\def\Tr{{\rm Tr}\, }

\begin{document}

\begin{flushright}
MIT-CTP-3143\\
DAMTP-2001-41\\
ITP-01-41 \\
ROM2F/2001/15\\
PUTP-1987\\
{\tt hep-th/0105276}
\end{flushright}

\begin{center}
{\Large\bf
How to go with an $RG$ Flow}
\vskip .8truecm

{\large\bf Massimo Bianchi,${}^{\star}${}\footnote{
{\tt M.Bianchi@damtp.cam.ac.uk}, on leave from {\it
Dipartimento di Fisica, Universit\`a di Roma ``Tor Vergata'', 00133
Rome Italy}} Daniel Z. Freedman${}^{\dagger}${}\footnote{
{\tt dzf@math.mit.edu}, on leave from {\it 
Department of Mathematics and Center for Theoretical Physics, MIT, 
Cambridge MA 02139, USA}} \\
and Kostas Skenderis${}^{\ddagger}$\footnote{{\tt
kostas@feynman.princeton.edu}}}\\
\vskip 0.5truecm
${}^{\star}$ {\it DAMTP, University of Cambridge \\
CMS, Wilberforce Road, Cambridge CB3 0WA, UK}
\vskip 0.5truemm
${}^{\dagger}$ {\it Institute for Theoretical Physics, 
University of California, 
\\
Santa Barbara, CA 93106, USA}
\vskip 0.5truemm
${}^{\ddagger}$ {\it Physics Department, Princeton University \\
Princeton, NJ 08544, USA}

\end{center}
\vskip .5truecm
\begin{abstract}
We apply the formalism of holographic renormalization to domain
wall solutions of 5-dimensional supergravity which are dual to
deformed conformal field theories in 4 dimensions.
We carefully compute one- and two-point functions of the 
energy-momentum tensor and the scalar operator mixing with 
it in two specific holographic flows, resolving previous difficulties with 
these correlation functions. As expected, two-point
functions have a 0-mass dilaton pole for the
Coulomb branch flow in which conformal symmetry is broken
spontaneously but not for the flow dual to a mass deformation in
which it is broken explicitly. A previous puzzle of the energy
scale in the Coulomb branch flow is explained.

\end{abstract}

\newpage
\section{Introduction}
\setcounter{equation}{0}

The holographic correspondence between (super)gravity theories on
$AdS$ spaces and (super)conformal field theories has passed many
tests and generated much insight into the strong coupling behavior
of field theory. For pure superconformal theories one can calculate
many correlation functions from 5-dimensional supergravity, a
procedure which is greatly facilitated by the high $SO(4,2)$ symmetry
of the bulk $AdS_5$ geometry. A number of domain wall solutions
of supergravity have also been found -- 5-dimensional geometries
with the symmetry of the 4-dimensional Poincar\'e group-- and general
arguments show that these are dual to $d=4$ superconformal theories
deformed either by addition of relevant operators to the Lagrangian
or by vacuum expectation values of such operators. For a Lagrangian
deformation,
conformal symmetry is explicitly broken and one expects that the trace
of the stress tensor $T_{ij}$ and the perturbing operator $\co$ are related
by $T^i_i(x) = \beta_{\co} \co(x)$, where $\beta_{\co}$ is the beta function
for the operator $\co$. For deformation by vacuum expectation value,
conformal symmetry is spontaneously broken and one expects that
$T^i_i =0$. Correlation functions of the stress tensor should thus be a
useful probe of the physics of holographic $RG$ flows, yet there is
a history of difficulty, briefly reviewed below, in attempts to calculate
correlation functions which display the expected physics from supergravity.

The purpose of this paper is to outline how these difficulties are
resolved by the formalism of holographic renormalization previously
developed for $AdS$ bulk geometry and various linear perturbations
in $AdS$ \cite{KSS}. This formalism embodies the duality between $UV$
divergences in the boundary field theory and $IR$ divergences of the on-shell
supergravity action. The $IR$ divergences are determined by near boundary
analysis of the classical supergravity equations of motion. The bulk
theory is then regularized by adding counterterms, expressed as
integrals of local expressions in the fields at boundary,
to cancel these divergences. This procedure
yields a finite renormalized action which is a functional of boundary
data for the bulk metric and other bulk fields. In the $AdS/CFT$
correspondence the boundary data are sources for dual operators and
$UV$ finite field theory correlation functions can be obtained by
functional differentiation with respect to these sources. These
correlators obey field theory Ward identities including conformal
anomalies.

Near boundary analysis is sufficient to resolve all divergences, but
leaves certain non-leading coefficients in the asymptotic expansion of
solutions undetermined. These coefficients contain full information on
the behavior of correlation functions at separated points in which
much of the physics resides. To find them one needs a full solution
of the equations of motion, usually specified uniquely by requiring
that the solution vanish in the deep interior of the bulk geometry.
A full solution of the nonlinear equations with general boundary data
is far too difficult, but one can linearize about the background
domain wall and, if fortunate, find  explicit fluctuations which
play the role of the bulk-to-boundary propagator in $AdS$ geometries.
Two-point correlation functions\footnote{In principle $n$-point correlators,
for $n \ge 3$ can be calculated through Witten diagrams, but the integrals
encountered are difficult in the reduced symmetry of domain walls. To our
knowledge they have never been attempted, and we respect this tradition.}
can be found using these fluctuations.
A simple cutoff method in momentum space \cite{gkp,fmrs} is sufficient
in many cases, but fails to give a full account of the stress tensor
correlators in $RG$ flows. We show how to determine these correlators
from the renormalized action. Our formalism also simplifies the calculation
of one-point functions.

Attention was first called to the stress tensor correlators in \cite{Notes}.
The linearized bulk field equations couple metric
fluctuations to those of the scalar fields which flow in the domain
wall backgrounds. It was shown how to decouple the equations for a
general domain wall, and
explicit fluctuations were found for two flows -- the $GPPZ$ flow \cite{gppz}
describing a supersymmetric mass deformation of $\cn =4$ SYM
 theory and
a Coulomb branch ($CB$) flow \cite{fgpw2,bs1} dual to spontaneous
breaking of the gauge
and conformal symmetry of  $\cn=4$ SYM by a vacuum expectation
value (vev) for the lowest chiral primary operator (CPO),
bilinear in the fundamental scalars.
Even with fluctuations known, the simple cutoff method failed
to give physically reasonable two-point correlators. Progress was made
in \cite{aft}. The main difference from \cite{Notes} was a different
choice of gauge for the bulk fluctuations which enabled the calculation
of the correlation function $\<T^i_i(x) T^j_j(y)\>$ for the $GPPZ$ flow.
A gauge invariant formulation of the problem is clearly desirable,
and this is incorporated in the method presented below.

There is other literature on fluctuations and correlation functions in
the flows just described, mostly for uncoupled fluctuations such as
transverse components of the bulk metric \cite{agpz,pstarinets, pz0002172},
or transverse bulk vectors dual to conserved currents \cite{bs2}.
The fluctuations of a number of other bulk fields, both coupled and
uncoupled, were found in \cite{Anatomy}, in which the implications of
supersymmetry and the multiplet structure of the fluctuations were
emphasized. Correlators in Coulomb branch solutions of 10-dimensional
Type $IIB$ supergravity were discussed in \cite{cheproib, giddings,
viswanathan}.

The flows we deal with are supersymmetric. They are dual to
boundary theories with unbroken $SUSY$, and $\<T_{ij}\>=0$ is
thus required. In field theory calculation with a supersymmetric
regulator automatically gives $\<T_{ij}\>=0$, but this relation would
generically fail with a non-supersymmetric regulator. We find an
analogous situation in our work in supergravity. The near boundary
analysis used in the holographic renormalization procedure
does not distinguish between $SUSY$ and non-$SUSY$ solutions of the
field equations, so manifest supersymmetry is not guaranteed. On
the other hand the renormalized action is ambiguous to the extent
that a restricted class of finite local counterterms can be added.
Requiring that the on-shell action evaluated on a bulk supersymmetric
solution vanishes selects such a finite counterterm and
$\<T_{ij}\>=0$ is then automatic.

Near boundary analysis of the field equations is straightforward but
quite complicated and
differs in detail from case to case depending on the dimension of the scalar
field in the flow and its potential in the bulk action. We therefore
try to be clear on the logical steps involved in the application
of the holographic renormalization method to RG flows. But we simply
present the asymptotic solutions of the field equations in an Appendix,
with details of the procedure to be
explained later \cite{later}.

The plan of the paper is as follows. In Sec 2 we review
the construction of supersymmetric domain walls in $D=5$. 
Our two examples can be lifted \cite{fgpw2,bs1,pw} to solutions
of $D=10$ Type $IIB$ supergravity. We argue that a Weyl transformation
should be made to a frame in which the $AdS_5$ scale becomes 
$L^2 = \a'$ (i.e. $L=1$ in string units). This facilates comparison
of energy scales in boundary and bulk and resolves a puzzle concerning
the size of the mass gap in the Coulomb branch solution of \cite{fgpw2,bs1}. 
We also outline how the holographic formalism leads to the definition
of the vevs $\<T_{ij}\>$ and $\<\co\>$ as functions of the sources. Higher
point correlation functions can be obtained from 
these quantities, and we derive the important Ward and trace identities
which they satisfy. In Sec 3 we explain how to use holographic
renormalization to determine  $\<T_{ij}\>$ and $\<\co\>$ explicitly in the
two flows we study. Physical vevs are obtained at this stage of the
program. In Sec 4 we give a general gauge invariant
treatment of the linear fluctuation equations. In Sec 5 we present
solutions of these equations and use our formalism to obtain two-point
correlation functions. Their physical properties are discussed.

\section{Holographic RG-flows}
\label{rgsection}
\setcounter{equation}{0}

Holographic RG-flows are described
by domain-wall spacetimes with scalar fields turned on. For one such
active scalar with canonical kinetic term,
the relevant part of the supergravity action is\footnote{
Our curvature conventions are as follows
$R_{\mu\nu\rho}{}^\sigma=\pa_\mu \G_{\nu\rho}{}^\sigma + 
\G_{\mu\lambda}{}^\sigma \G_{\nu\rho}{}^\lambda - \mu
\leftrightarrow \nu$ and $R_{\mu\nu}=R_{\mu\lambda\nu}{}^\lambda$. They differ
by an overall sign from the conventions in \cite{Notes,Anatomy}.}
\be
S = \int_M \de^5x \sqrt{G} \left[{1 \over 4} R
+\half G^{\m \n} \pa_\m \F \pa_\n \F + V(\F) \right]
-\half \int_{\pa M} \sqrt{\c} K
\label{scaction}
\ee
where $K$ is the trace of the second fundamental form.
We work in Euclidean signature.

We are interested in domain wall solutions of the equations of motion
(EOM's) resulting from
(\ref{scaction}) that preserve 4D Poincar\'e invariance. For one convenient
choice of radial coordinate, they take the form
\bea
\label{dmw}
ds^2&=&e^{2 A(r)} \d_{ij} dx^i dx^j + dr^2  \nonu
\Phi&=& \Phi(r)
\eea

The specific flows we consider are supersymmetric. From Killing spinor
conditions in the bulk supergravity theory one can deduce the
first order flow equations \cite{FGPW}
\be
\label{1storder}
{dA(r)\over dr} =- {2 \over 3} W(\F), \qquad {d \F (r)\over dr} =
\pa_\F W (\F)
\ee
where $W(\F)$ is the superpotential. The potential $V(\F)$ is expressed
in terms of $W$ by
\be \label{potential}
V(\F)={1 \over 2} (\pa_\F W)^2 - {4 \over 3} W^2 .
\ee
It is usually straightforward to solve the first order equations and
any such solution automatically satisfies the second order equations
of (\ref{scaction}) for domain walls (but not conversely).

Supersymmetry guarantees that the domain wall solution is stable.
However, the first order system can be derived from the requirement
of gravitational stability of asymptotically $AdS$ geometries even when the
action (\ref{scaction}) is not the truncation of a bulk supergravity
theory. A generalized positive energy argument \cite{posthm},
was used in \cite{townsend}, and it was shown that
$V(\F)$ must have the form (\ref{potential}) when there is a single
scalar field\footnote{For several scalars the obvious generalization
of the form (\ref{potential}) implies stability, but the converse
is not necesarily true.}. The argument in \cite{townsend}
implies that there is a ``superpotential'' $W$ such that
the critical point of $V(\F)$ associated with the $AdS$ geometry
is also a critical point of $W$.\footnote{
If one relaxes this  requirement the potential
can always be written in the form (\ref{potential}); one just
views (\ref{potential}) as a differential equation for $W$ \cite{DFGK}.
In this case, however, as the original critical point may not be
a critical point of $W$, the results of \cite{townsend}
about gravitational stability do not necesarily apply.}
In the $AdS/CFT$ correspondence, positivity of energy about a
given $AdS$ critical point is mapped into unitarity of the
corresponding CFT. It follows that in all cases the dual
CFT is unitary,  the potential can be written as
in (\ref{potential}) \cite{ST}.
When (\ref{potential}) holds a simple $BPS$ analysis \cite{ST,DFGK}
of the domain wall action yields the flow equations (\ref{1storder}).

We assume that $W(\F)$ has a stationary point at $\F=0$. Near this point
$W(\F)$ can be approximated by
\be
\label{approxW}
W(\F) \approx - \left[
{3 \over 2} + {\mu \over 2} \F^2 + \co(\F^3) \right]
\ee
and we assume that $0 < \mu < 4$. For large positive $r$ the domain wall
solution is well approximated by the boundary region of an $AdS_5$
geometry with scale $L=1$, i.e.
\be
A(r) \approx r \qquad  \F \approx \exp(- \mu r)
\ee
If $0< \mu <2$, then the bulk field $\F$ is dual to an operator $\co$
of dimension $\Delta = 4 - \mu$ and the domain wall describes a
relevant deformation of the CFT Lagrangian. If $ 2 \leq \mu < 4$
then the dual operator has scale dimension $\Delta = \mu$ and the
bulk flow describes spontaneous breaking by the vev $\<\co\>$.\footnote
{For $\mu=\Delta=2$ the scalar solution dual to a Lagragian deformation
is $\F \approx r \exp(-2r)$. Such a purely bosonic mass
deformation breaks $SUSY$. Nevertheless one can obtain such solutions
from the first order equations. The relevant ``superpotential''
is given by
$W(\F(r))= -3/2 - \exp(-4 r)(1/8 - r/2 + r^2)$\ .
This ``superpotential'' is not analytic in $\F$
(to obtain $W(\F)$ one needs to invert $\F = r \exp(-2r)$).
}\ {}\footnote{A more detailed
argument \cite{KleWit} shows that
operators of dimension $1 \leq \mu <2$ can be described holographically.}

In the discussion above we considered domain-wall solutions that
asymptote to $AdS_5$ spacetimes with scale $L=1$, rather than the
official $AdS/CFT$ scale $L = (4 \p \a'^2 g_s N)^{1/4}$. 
This is a significant point that we now discuss in some detail.
As we review in the next sub-section, the on-shell gravitational action
evaluated on the near-horizon solution is equated
to the gauge theory effective action by the $AdS/CFT$ correspondence.
The latter has a large $N$ expansion of the form
\be 
\label{gauge}
S_{gauge} = \sum_{g=0}^\infty N^{2-2g}
\sum_{k=0}^\infty c_{g,k} (g_{YM}^2 N)^k
\ee
On the gravitational side, specifically $IIB$ string theory in the usual
Einstein frame, the overall constant in front
of the action is $1/16\pi G_N^{(10)}$, where
\be \label{GN1}
G_N^{(10)} = 8 \pi^6 g_s^2 \a'^4
\ee
is the $D=10$ Newton's constant.
So the gravitational 
action seems at first sight to have a different leading behavior.
As is well known the correct dependence  on $N$ and $4 \pi g_s=g_{YM}^2$
is restored because the near-horizon $D3$-brane metric depends on
these quantities. To provide a more manifest match to the gauge theory
expansion we perform a constant Weyl rescaling \cite{BST}
so that the solution now becomes asymptotic to $AdS_5 \times S^5$
with $AdS$ and sphere radius equal to one in string units, i.e. $L^2 = \a'$.
Newton's constant is now equal to
\be \label{GN2}
G_N^{(10)} = 8 \pi^6
{\a'^4 \over 16 \pi^2 N^2}  = 8 \pi^6 g_s^2 
\left({\a' \over \sqrt{4\pi g_s N}}\right)^4
\ee
The five dimensional Newton's constant $G_N^{(5)}$ is now obtained by dividing
(\ref{GN2}) by the volume of the unit five-sphere, $vol(S^5)=\p^3$.
It follows that the overall constant in front of the five dimensional 
action (\ref{scaction}) is $N^2/2 \p^2$.
The difference between (\ref{GN1}) and (\ref{GN2}) means that
energies are measured in different units, and the effective $\a'$ in the two 
frames differ by a factor of $\sqrt{g_s N}$. 
This is the origin  of the factor $\sqrt{g_s N}$
present in the UV/IR relation derived in \cite{PP}.

The Weyl scaling needed is the special case $(p=3)$ of the
rescaling used
to reach the so-called ``dual-frame'' \cite{BST}\footnote{
The dual-$Dp$-frame is defined as the frame where the metric
and the $8-p$-field strength couple to the dilaton the
same way.}. In this frame
all $Dp$-branes\footnote{Fivebranes are exceptional in that    
the near-horizon limit gives $Mink_7 \times S^3$ with a linear 
dilaton.}, not just the $D3$-brane, 
have a near-horizon limit $AdS_{p+2} \times S^{8-p}$. 
The Weyl rescaling and the dilaton are not
constant when $p \neq 3$ and because of this 
the solutions are 1/2 supersymmetric
even in the near-horizon limit.
It has been argued in \cite{BST}
that this frame is holographic. It is in this frame that
the on-shell gravitational action has manifestly the same leading
behavior at large $N$ as the gauge theory effective action,
and the supergravity variables automatically take into
account the UV/IR relation.
So in order to have manifest matching between gauge theory
and supergravity computations (without worrying about different units)
we need to do the gravitational computation in the dual-frame.
For the case at hand this simply means that we will consider the
solution with $L=1$ (we set $\a'=1$ from now on), and Newton's
constant is equal to $1/N^2$.

\subsection{Correlators and Ward Identities}

To obtain correlation functions we must go back to the second order
EOM's of (\ref{scaction}) and consider solutions with arbitrary Dirichlet
boundary conditions for the bulk fields. As in earlier papers on
holographic renormalization, we use a different radial variable
$\rho = \exp (-2r)$, where the boundary is mapped to $\rho =0$.
The bulk scalar satisfies the (modified) Dirichlet condition
\be
\F(\rho,x) \rightarrow  \rho^{{4-\Delta \over 2}} \phi_{(0)}(x) \quad
({\rm as} \ \rho \rightarrow 0)
\ee
The general bulk metric ansatz is
\be \label{coord}
ds^2=G_{\m \n} dx^\m dx^\n = {d\r^2 \over 4 \r^2} +
{1 \over \r} g_{ij}(x,\r) dx^i dx^j
\ee
where the boundary metric limits to
\be \label{glim}
 g_{ij}(x,\r) \rightarrow g_{(0)ij}(x) \quad ({\rm as} \ \r \rightarrow 0)
\ee

In the $AdS/CFT$ correspondence, the boundary data $ g_{(0)ij}(x)$
and $\phi_{(0)}(x)$ are arbitrary functions of the transverse
coordinates $x^i$ and are the sources for the stress tensor $T_{ij}$
and, respectively, the operator $\co$ in the boundary field theory.
The correspondence is then expressed by the basic formula for the
generating functional
\be \label{basic}
\< \exp(-S_{QFT}[g_{(0)}] - \int d^4x \sqrt{g_{(0)}} \co(x)
\phi_{(0)}(x))\>
= \exp(-S_{SG}[g_{(0)},\phi_{(0)}])
\ee
On the left side $\<...\>$ denotes the functional integral average involving
the field theory action $S_{QFT}[g_{(0)}]$ minimally
coupled to $g_{(0)}$.
The action  $S_{SG}[g_{(0)},\f_{(0)}]$ is the classical action integral of
(\ref{scaction})
evaluated on the classical solution with boundary data $g_{(0)}$ and
$ \phi_{(0)}$. This is the leading term in the semiclassical computation
of the supergravity partition function.
Unless confusion arises we will henceforth use $S$ to
denote this classical on-shell action.

Actually $S$ is divergent due to the behavior of the solution $g_{ij}(\r,x),
\F(\r,x)$ near the $AdS$ boundary. One must regularize the action, e.g.
by cutting off the radial integration at a small value $\r=\epsilon$,
and add appropriate boundary
counterterms $S_{{\rm ct}}$ to cancel the divergences.
The action $S$ must then be
replaced by the renormalized
$S_{{\rm ren}}= \lim_{\e \to 0}(S_{{\rm reg}}+S_{{\rm ct}})$
in (\ref{basic}) and $S_{{\rm ren}}$
becomes the generating functional of connected correlation functions.
This process is the heart and soul of holographic renormalization and is
described in the next section. In the rest of this section we discuss
some general properties of the correlation functions obtained from the
procedure.

By definition, the variation of the effective action
is equal to\footnote{Notice that the definition of $\< \co \>$ 
differs by a sign from the definition used in \cite{KSS}.}
\be \label{variation}
\delta S_{{\sm ren}}[g_{(0) ij}, \f_{(0)}] =
\int d^4x \sqrt{g_{(0)}} [\half \<T_{ij}\> \delta g_{(0)}^{ij}
+ \<\co \> \delta \f_{(0)}]
\ee
The expectation values $\<T_{ij}\>$ and $\<\co\>$ are functions of the
sources to be computed in the next section.
Multi-point correlation functions can be obtained by further
differentiation, e.g.
\be
\< T_{ij}(x)T_{kl}(y) \> = - {2 \over \sqrt{g_{(0)}(y)}}
{\d \< T_{ij}(x) \> \over
\d g_{(0)}^{kl}(y)} .
\ee

A well known feature of $AdS/CFT$ physics is the correspondence between
bulk gauge symmetries and global symmetries of the boundary theory. In
the present setting the relevant gauge symmetries are bulk diffeomorphisms.
Some of these were used to bring the bulk metric to the form (\ref{coord}).
The remainder preserve the form (\ref{coord}), and  we distinguish between
diffeomorphisms involving only the 4 transverse coordinates:
\be \label{diff4}
\delta g_{(0)}^{ij} = - (\nabla^i \xi^j + \nabla^j \xi^i), \qquad
\delta \f_{(0)} = \x^i \nabla_i \f_{(0)}
\ee
and a one-parameter subalgebra of $5D$ diffeos \cite{schwimmer} whose
effect on the boundary data coincides with the Weyl transformation
\be \label{weyl}
\delta g_{(0)}^{ij} = - 2 \s g_{(0)}^{ij}, \qquad
\delta \f_{(0)}=(\D - 4) \s \f_{(0)}
\ee
The counterterms needed to make the on-shell action finite
are manifestly invariant under the $4D$ diffeomorphisms,
but they break the diffeos that induce Weyl
transformations  on the boundary \cite{HS}.
It follows that the diffeomorphism Ward identity should
hold, but the trace Ward identity is expected to be broken.

It is straightforward to substitute the variations (\ref{diff4})-
(\ref{weyl}) in
(\ref{variation}) and obtain the Ward and trace identities\footnote{\label{QFT}
Since $\<T_{ij}\>$ includes the scalar source term in (\ref{basic})
these identities
do not quite have the standard field theory form. For this one must use
$\<T_{ij}\>_{QFT} = \<T_{ij}\> + \f_{(0)} \<\co\> g_{(0)ij}.$} 
\be
\nabla^i \<T_{ij}\> = -  \<\co\> \nabla_j \f_{(0)} \label{divW}
\ee
\be \label{trace}
\<T^i_i\>= (\Delta - 4) \f_{(0)} \< \co \> + \ca
\ee
where $\ca$ is the conformal anomaly. As mentioned above, it
originates from the fact that the counterterms break part
of the $5D$ diffeomorphisms. The explicit form of $\ca$ will be
determined below. It is useful to note here that $\ca$ is local
in the sources. One of the most elegant aspects of the holographic
renormalization formalism is the simple emergence of Ward and trace
identities including conformal anomalies.

The source for the stress energy tensor can be decomposed
as follows,
\be \label{deltag}
\d g_{(0) ij} = \d h^{T}_{(0)ij} + \nabla_{(i} \d h^{L}_{(0)j)}
+ \d_{ij} {1 \over 4} \d h_{(0)}
- \nabla_{i} \nabla_{j} \d H_{(0)}
\ee
where
\be
\nabla^i h^{T}_{(0)ij}=0,\ \ h^{T\ i}_{(0)i}=0, \qquad
\nabla^i h^{L}_{(0)i}=0 \ .
\ee
All covariant derivatives are that of $g_{(0)}$.

Using (\ref{deltag}) and (\ref{divW})-(\ref{trace}) and partial integration
we can rewrite (\ref{variation}) as
\bea
\d S_{\rm ren}&=&\int \de^4x \sqrt{g_{(0)}}\left(
-\half \d h^{T\ ij}_{(0)}  \<T_{ij}\>
- \half \d h^{L\ i}_{(0)}  \< \co \> \nabla_i \f_{(0)} \right. \nonu
&&
- \half
\d H_{(0)} [\nabla^i \<O\> \nabla_i \f_{(0)} + \<\co \>
\nabla^2 \f_{(0)}] \nonu
&&\left.
-{1 \over 8} \d h_{(0)} [(\Delta - 4) \f_{(0)} \< \co \> + \ca]
+\d \f_{(0)} \< \co \>\right)
\eea
The form above is quite general, but we now make two assumptions appropriate
to our situation. We assume that the sources describe an $x$-independent
domain wall and linear fluctuations above it. This is sufficient to
study one- and two-point functions.
In this case  either $\nabla^i \<\co\>_B=0$ or $\nabla_i \f_{(0)B}=0$,
where the sub-index denotes  a background value.
The term proportional to $\d h^{L\ i}_{(0)}$ vanishes because
either the source or the vev is constant and in the latter case the
term drops out upon partial integration. 
One then finds
\bea \label{finalS}
\d S_{{\rm ren}}&=&\int \de^4x \sqrt{g_{(0)}}[
-\half \d h^{T\ ij}_{(0)}  \<T_{ij}\> \\
&& - \half \d H_{(0)} \<\co\> \nabla^2 \f_{(0)}
 +{1 \over 8} \d h_{(0)}[(4-\Delta) \f_{(0)} \< \co \> -\ca ]
+\d \f_{(0)} \< \co \> ] \nonumber
\eea
Since $\<T_i^i\> = -8 \d S_{{\rm ren}}/\d h_{(0)}$ the
expression in (\ref{finalS}) shows that
all correlation functions of  $T^i_i$ and $\co$ can be obtained from
the form of $\<\co\>$ and $\ca$ as functions of the sources. This is a
consequence of the trace identity (\ref{trace}).

The two-point function of $T_{ij}$ has the standard representation
\be
\<T_{ij}(p)T_{kl}(-p)\> =\P^{TT}_{ijkl} A(p^2) +\pi_{ij} \pi_{kl} B(p^2)
\ee
in terms of the projection operators  $\pi_{ij} =\delta_{ij} -p_ip_j/p^2$
and
\be
\P^{TT}_{ijkl} \equiv - {\d h^T_{(0)ij} \over \d h^{T\ kl}_{(0)}} =
{1 \over 2} (\p_{ik} \p_{jl} + \p_{il} \p_{jk})-{1 \over 3} \p_{ij} \p_{kl}
  \quad .
\ee
The transverse traceless $(TT)$ amplitude $A(p^2)$ can thus be calculated
by further variation of the $TT$ projection of $\<T_{ij}\>$, while
$B(p^2) = \<T^i_i(p) T^j_j (-p)\>/9.$ The Ward identity implies that
$\<T_{ij}(p) \co (-p)\> = \pi_{ij} C(p^2)$ with invariant amplitude 
$C(p^2)= \<T^i_i(p) \co (-p)\>/3$. Note that $\<T_{ij}(p) \co (-p)\>$ 
is the connected 
correlator. When the background vev $\<\co\>$ does not vanish 
one must correct for the
source term in Footnote \ref{QFT} to obtain this. We will obtain 
these correlators, together
with $\<\co (p) \co (-p)\>$ for the $CB$ and $GPPZ$ flows in Sec. 5.

\section{Holographic Renormalization}
\setcounter{equation}{0}

We will use the renormalization method pioneered in \cite{HS} and
developed in \cite{KSS}, see also \cite{BK} for related work.
In this method one regulates the divergent on shell action $S$ by
restricting the $\rho$ integration to $\rho \geq \e$.
Asymptotic solutions of the field equations with arbitrary
Dirichlet boundary data (= field theory sources)
are then obtained and used to
express the divergences as  $\e \rightarrow 0$ in terms of the sources.
Finally one adds  counterterms to cancel these divergences.
Here we will only give the results; details will be presented elsewhere.

The solutions we consider are asymptotically $AdS$. This means that
near the boundary one can find coordinates such that the metric
$g_{ij}(x,\r)$ in (\ref{coord}) can be expanded in a series of the form
\be \label{series}
g(x,\r)=g_{(0)} + g_{(2)} \r + \r^{2} [g_{(4)}
+ h_{1(4)} \log \r + h_{2 (4)} \log^2 \r] + ...
\ee
The scalar field can also be expanded in a similar
fashion.  Since the exact form of the expansion depends on the
mass of the bulk scalar,  it will be presented below
for the two cases we consider.

The next step is the near boundary analysis of the EOM's. In this
process one subtitutes the assumed expansions into the EOM's and solves
them iteratively. In this way many higher order terms in the
expansion are determined as (local) functions of the sources, but not all
terms are so determined. For the metric, $g_{(4)}$ is the first term
which is not fully determined (although its trace and covariant
divergence are determined). It is to be expected that near boundary
analysis does not completely fix the solution of second order field
equations with a Dirichlet condition on an $AdS$ boundary. Additional
information on the behavior in the deep interior of the space-time is
required. The product of this phase of the procedure
is an asymptotic solution of the EOM's in which the unspecified
coefficients are simply carried as such within the series expansions.

It turns out that divergences of the cutoff action are fully
determined in terms of the sources by near boundary analysis. The
divergences can be expressed as counter terms which are
boundary integrals of local invariants constructed from the induced metric
$\g_{ij}= {1 \over \e} g_{(0)ij}(x)$ and the scalar field $\F(x,\e)$.
One adds these counter terms to define the finite renormalized action
$S_{{\rm ren}}
= S_{{\rm reg}} + S_{{\rm ct}}$ in which the limit
$\e \rightarrow 0$ may be
taken.

The exact form of the counterterms depends on the specific
potential of the scalar field. For a given potential, however,
the derived counterterms are universal, i.e. the on-shell action
will be finite for {\em any} solution of the bulk field equations.
This is a property that any holographic renormalization scheme
should have.

The renormalization  procedure is ambiguous
to the extent that finite local counter terms can be added to
$S_{{\rm ren}}$. This corresponds to scheme dependence
in quantum field theory. In particular, $S_{{\rm ren}}$ need not incorporate
requirements of supersymmetry, since it was derived using counterterms
valid for the most general non-supersymmetric solution of the EOM's.
The particular requirement which need not be satisfied is that
$E_{vac} =0$ in a supersymmetric vacuum. In holography this means
that $S_{{\rm ren}} =0$ when evaluated in the background geometry of
a solution of (\ref{1storder}). We will use a
supersymmetric scheme to fix the form of the counterterms.

It is much easier to compute the regularized on-shell action for domain
wall solutions
(\ref{dmw}) than for those with $x$-dependent boundary data.
For solutions of (\ref{1storder}) the answer can be
read off from the $BPS$ form of the action in \cite{ST,DFGK} (see (13)
in \cite{ST} or (14) in \cite{DFGK}). One obtains
\be \label{bkgd}
S_{{\rm bkgd,reg}} = \int_{\r=\e} d^4 x \sqrt{\gamma} W[\Phi]
\ee
where $\gamma_{ij} = e^{2 A} \d_{ij}$ is the induced metric
at the cutoff surface $\r=\e$ (or equivalently
$r=-{1 \over 2} \log \e$). When $W(\F)$ is expressed as a series, as in
(\ref{approxW}), the low order terms will be divergent and these must
agree with similar terms obtained from the divergences of more
general solutions. In addition, depending on the asymptotic behaviour
in $\rho$ of the scalar field, there
may be a residual finite part. If present this must be subtracted for
the scheme to be supersymmetric, ensuring that $S_{{\rm ren}}=0$ in the
background. The specific working of this mechanism will be discussed
with our examples below.

It is both interesting and helpful that one can determine some divergences
of the general $S_{\rm reg}$ by examining simple subclasses of solutions of
the theory, in our case SUSY domain wall solutions. However, one must
study more general solutions in order to obtain conterterms necessary
to cancel divergences in all correlation functions. For example, there
are counterterms involving $\partial_i \phi_{(0)}$ and
$\partial_i g_{(0)jk}$, which cannot be found using $x$-independent
domain wall solutions.

\subsection{Coulomb branch}

Our first example is the supergravity dual of a particular state in the
Coulomb branch of $\cn = 4$ SYM theory. It can be obtained by turning
on the $SO(4) \times SO(2)$ singlet component of the scalar field dual
to CPO in the ${\bf 20'}$ of $SO(6)$. We henceforth denote the active
scalar field by $\Phi$.

The superpotential is given by
\be
W(\Phi)=-e^{-{2 \F \over \sqrt{6}}} - \half e^{{4 \F \over \sqrt{6}}}
      = -{3 \over 2} - \F^2 + \co(\F^3)
\ee
The domain-wall solution is given by
\be
v=e^{\sqrt{6} \Phi}, \qquad e^{2A}=l^2 {v^{2/3} \over 1-v}, \qquad
{d v \over d r} = 2 v^{2/3} (1-v)
\ee
The boundary is at $v=1$. There is a curvature singularity at $v=0$ which
is the origin of a disc distribution of $D3$-branes in Type $IIB$
supergravity. See \cite{fgpw2,bs1,Anatomy,bs2} for more details of this
solution and previously studied correlation functions.

The change of variables
that brings the domain-wall metric to the coordinate system (\ref{coord})
is given by
\be
1-v=l^2 \r -{2 \over 3} l^4 \r^2+ \co(\r^3)
\ee
In these coordinates the solution $\vf_B, A$,
is given by
\be \label{CBsolution}
\vf_B={1 \over \sqrt{6}}(-\rho l^2 + {1 \over 6} l^4 \r^2 + \co(\r^3)), \ \
e^{2 A} = {1 \over \r}(1 -{1 \over 18} l^4 \r^2 + \co(\r^3))
\ee

By inspection (\ref{bkgd}) evaluated on the background solution
(\ref{CBsolution}) is divergent. It follows
from the given asymptotics that only the first two terms in the
expansion of $W$ around $\F=0$ contribute to the IR divergences,
and that there is no finite term.
All the other terms in the expansion of $W$ vanish in the limit $\e \to 0$.
Thus the counterterms needed to make the background action
finite are given by
\be \label{ctbkgd}
S_{{\rm ct,bkgd}} = \int_{\r=\e} \de^4 x \sqrt{\g}
\left({3 \over 2} + \F^2\right)
\ee
We will see that these are part of the counterterms required to
make finite the on-shell action in general.
Since there is no finite term left after the subtraction,
the renormalized action evaluated on the background solution
equals zero, as required by supersymmetry.

To obtain the general form of the counterterms we note that
the asymptotic expansion for a bulk scalar field of $AdS$ mass $m^2=-4$
dual to an operator of conformal dimension $\D=2$ reads,
\be \label{expCB}
\F(x,\r)=\r \log \r (\f_{(0)} + \f_{(2)} \r + \r \log \r \psi_{(2)} + ...)
+ \r (\tf_{(0)} + ...)
\ee
Inserting the asymptotic expansions in the bulk field
equations one finds that the
coefficients shown in (\ref{series}) are uniquely determined
in terms of $g_{(0)}$ and $\f_{(0)}$ except for $g_{(4)ij}$ that is
only partially determined and $\tf_{(0)}$ that is undetermined.
In particular, only $\Tr g_{(4)}$ and $\nabla^i g_{(4)ij}$ are determined.
The exact expressions are given in the appendix.
The coefficients $g_{(4)ij}$ and $\tf_{(0)}$
are related to the holographic one-point functions
in the presence of sources \cite{KSS} as we will derive shortly.

Knowledge of the asymptotic solution allows one to evaluate
the regularized action and obtain the divergences.
These can be cancelled by adding the following covariant
counterterms
\bea
S_{\sm{ct}}&=&\int_{\r=\e} \de^4 x \sqrt{\g} \left(
\left[{3 \over 2}- {1 \over 8} R
-{1 \over 32} \log \e  (R_{ij} R^{ij} - {1 \over 3} R^2)\right] \right.\nonu
&&\left.\qquad \qquad \qquad
+ \left[\Phi^2(x,\e)  + {\F^2(x,\e) \over \log \e}\right] \right),
\eea
where $\c$ is the induced metric at $\r=\e$.
All curvatures are of the induced metric. Notice that
the term $\F^2/\log \e$ is divergent in the limit $\e \to 0$
with $\phi_{(0)}$ fixed (but equal to zero when $\f_{(0)}=0$
which is the case for the SUSY domain wall solutions).
Thus the set of counterterms contains (\ref{ctbkgd}) and more.

The renormalized action is equal to
\be \label{ren}
S_{\sm{ren}}=\lim_{\e \to 0}
\left(S_{\sm{reg}} + S_{\sm{ct}} \right)
\ee
where $S_{\sm{reg}}$ is the on-shell action in (\ref{scaction})
regulated by restricting the range of integration to $\r \geq \e$.

The expectation value of the operator dual to $\Phi$ is given by
\be
\< \co \> = {1 \over \sqrt{g_{(0)}}}
{\delta S_{\sm{ren}} \over \delta \f_{(0)}} =
\lim_{\e \to 0} \left( {\log \e \over \e} {1 \over \sqrt{\g}}
{\delta S_{\sm{ren}} \over \delta \F(x,\e)} \right)
\ee
By straightforward computation of the variational derivative \cite{later}
one obtains
\be \label{oexpe}
\< \co \> = 2 \tf_{(0)}.
\ee

One can similarly compute \cite{later} the expectation value
of the stress-energy tensor. By definition
\be \label{tij1}
\<T_{ij}\> = {2 \over \sqrt{g_{(0)}}}
{\d S_{\sm{ren}} \over \d g_{(0)}^{ij}}
=\lim_{\e \to 0} \left({1 \over \e}
{2 \over \sqrt{\c(x, \e)}} {\d S_{\sm{ren}} \over \d \c^{ij}(x,\e)}
\right)
\ee
After some computation one finds that all infinities cancel
and the finite part is equal to
\bea \label{stress}
\<T_{ij}\>&=&
g_{(4)ij} +
{1 \over 8}[\Tr\, g_{(2)}^2 - (\Tr\, g_{(2)})^2]\, g_{(0) ij}
- \half (g_{(2)}^2)_{ij}  \\
&+& {1 \over 4}\, g_{(2) ij}\, \Tr\, g_{(2)}
+{1 \over 3} (\tf_{(0)}^2 -3 \f_{(0)} \tf_{(0)}) g_{(0)ij}
+ {2 \over 3} \f_{(0)}^2 g_{(0)ij} + {3 \over 2} h_{(4) ij} \nonumber
\eea
The last two terms can be cancelled by adding finite local counterterms
to the action. The last term is proportional to the stress energy 
tensor derived from an action given by  the gravitational 
conformal anomaly (\ref{gravanom}) \cite{KSS}.
The next-to-last term is proportional to the stress energy 
tensor derived from an action equal to the matter
conformal anomaly (\ref{cbscalanom}). 

The computation of $\< \co \>$ and $\<T_{ij}\>$ was carried out 
in a specific coordinate system. One may wonder whether 
these results are sensitive to our specific choice. From the
derivation of $\< \co \>$ and $\<T_{ij}\>$ as functional
derivatives of $S_{{\rm ren}}$ it follows that they transform
as tensors up to the contribution of the conformal anomaly.
The contribution of the conformal anomaly to the transformation
rules is also straightfoward to obtain in all generality \cite{KSS,strings}. 
Here we only discuss how 
$\< \co \>$ and $\<T_{ij}\>$ transform under a constant 
rescaling of the $\r$ variable. This is of particular interest 
because the much-discussed association of the bulk radial 
coordinate with energy scale in the boundary field theory
suggests that the transformation 
\be \label{rresc}
\r = \r' \mu^2, \qquad 
x^i= x^i{}' \mu
\ee
introduces the RG scale $\mu$. This transformation is an isometry of 
$AdS$ spacetime (with a metric given by (\ref{coord}) with 
$g_{(0)ij}=\d_{ij}$ and all other $g_{(k)ij}=0$). Under this transformation
most coefficients pick up overall factors of $\mu$ according to 
their dimension, but there are also non-trivial transformations 
due to the logarithms in the asympotic solutions. 
One obtains
\bea
\f_{(0)}'(x') &=& \mu^2 \f_{(0)}(x' \mu), \qquad
g_{(0)}'(x') = g_{(0)}(x' \mu),
\qquad g_{(2)}'(x')=\mu^2 g_{(2)}(x' \mu) \nonu
\tf_{(0)}'(x')&=& \mu^2 [\tf_{(0)} (x' \mu)+ \log \mu^2 \f_{(0)}(x' \mu)], \\
g_{(4)}'(x')&=& \mu^4 [g_{(4)} 
+ \log \mu^2 (h_{(4)} - {2 \over 3} \f_{(0)} \tf_{(0)} g_{(0)}) 
- (\log \mu^2)^2 {1 \over 3} \f_{(0)}^2 g_{(0)}](x' \mu)
 \nonumber
\eea
It follows that 
\bea
&&\< \co(x') \>'= \mu^2 \left(\< \co(x' \mu) \> 
+ \log \mu^2 2 \f_{(0)}(x' \mu)\right) \\
&&\< T_{ij}(x') \>' = \mu^4 \left(\< T_{ij}(x' \mu) \> 
+ \log \mu^2 [h_{(4)ij} 
- \f_{(0)}^2 g_{(0)ij}](x' \mu) \right) \nonumber
\eea
It is satisfying to see that the new terms can be obtained 
from the  following local finite counterterm,
\be \label{finanom}
S_{{\rm fin}}(\mu) = \int \de^4x \sqrt{g_{(0)}} \log \mu^2 \half \ca
\ee
where $\ca=\ca_{grav}+ \ca_{scal}$ is the conformal anomaly given in 
(\ref{gravanom}) and (\ref{cbscalanom}). This implies the transformation
(\ref{rresc}) only adds contact terms to correlation functions,
and scales the momenta by $1/\mu$.

{}From the expressions in (\ref{oexpe}) and (\ref{stress})
we find the vevs with all sources equal to zero
are given by
\be
\<\co\>_B=-{2 \over \sqrt{6}} {N^2 \over 2 \p^2} l^2, \qquad \<T_{ij}\>_B =0
\ee
The term $N^2/2 \p^2$ is the overall constant in front of the 
action\footnote{Here and henceforth we adopt the policy of including this
factor only when final results for correlation functions are given.}
discussed in section 2. The vevs 
show that the solution describes a state in the
moduli space of vacua, as promised. As one might have expected
the size of the vev, $l^2$, is set by the size of the symmetry
breaking effect; the radius of the disk distribution of
D3-branes.

It is straightforward to use
the solution of the bulk field equations given in
appendix A.1 to show that
\bea
\nabla^i \<T_{ij}\> &=& -  \<\co\> \nabla_j \f_{(0)} \label{divWCB} \\
\<T^i_i\>&=&- 2 \f_{(0)} \< \co \>
+{1 \over 16} (R_{ij} R^{ij} -{1 \over 3} R^2)
+ 2\phi_{(0)}^2  \qquad \label{trWCB}
\eea
i.e. $ \<T_{ij}\>$ correctly
satisfies the diffeomorphism and trace Ward identities.
The last two terms in (\ref{trWCB}) are what we called $\ca$ in
(\ref{trace}). The second term
\be
{\ca}_{grav} = {1 \over 16} (R_{ij} R^{ij} -{1 \over 3} R^2)
\label{gravanom}
\ee
is the holographic gravitational
conformal anomaly \cite{HS} and the last term
\be
{\ca}_{scal} = 2\phi_{(0)}^2
\label{cbscalanom}
\ee
is the conformal
anomaly due to matter \cite{PeSk}. The coefficients in both
of them are known not to renormalize, and indeed we obtain
the correct value. The Ward identities and the anomalies
are important checks of the intermediate computations and of the
consistency of the formalism.

\subsection{$GPPZ$ flow}

Our second example is the supergravity dual of a $\cn = 1$
supersymmetry preserving mass deformation of $\cn = 4$ SYM theory
\cite{gppz}. We will consider only the simplest case in which the
active scalar field is one of the two $SO(3)$ singlet, dimension
$\Delta=3$ scalars studied in \cite{gppz}. Specifically we
consider the field called $m$, here renamed $\F$, which is dual
to a chiral fermion mass operator, and we do not treat a more general
flow involving $m$ and the second scalar $\sigma$.

The superpotential reads
\be \label{GPPZW}
W(\F) = -{3\over 4} \left[ 1 + \cosh \left(2 \F \over \sqrt{3}\right) \right]
      = -{3 \over 2} - \half \F^2 - {1 \over 18} \F^4 + \co(\F^5)
\ee
The domain-wall solution is given by
\be
\vf_B = {\sqrt{3}\over 2} \log {1 +\sqrt{1-u} \over 1-\sqrt{1-u}},
\qquad e^{2 A} = {u \over 1-u}, \qquad {d u \over dr} = 2 (1-u).
\ee
The $u$ variable is related to the $\r$ variable by $u=1-\r$.
Near the boundary the solution has the expansion
\be \label{solrho}
\vf_B = \r^{1/2} [\sqrt{3} + \rho {1 \over \sqrt{3}} + \co(\r^2)], \qquad
e^{2 A}={1 \over \r}(1-\r) .
\ee
By inspection one finds that the on-shell action (\ref{bkgd})
evaluated on this background is divergent, and that only the
first two terms in the expansion of $W$ contribute to the
IR divergences. To cancel them we add the counterterms
\be \label{ctbkG}
S_{{\rm ct,bkgd}} = \int \de^4x \sqrt{\g} \left[{3 \over 2} +
\half \F^2\right]
\ee
With the addition of these terms the on-shell action is finite, but
not zero because the $\F^4$ term has a finite limit. The subtraction
of finite terms corresponds to a choice of scheme. We proceed by subtracting
the finite term so that the on-shell value of the action is zero
when evaluated on the background, as required by supersymmetry.
In other words, we supplement the counterterms in (\ref{ctbkG})
by the finite counterterm
\be \label{ctfinite}
S_{{\rm ct,fin}}=\int d^4x\sqrt{\g} {1 \over 18} \F^4
\ee

We now proceed to obtain the general counterterms.
For a scalar of $AdS$ mass $m^2=-3$, dual to an operator of
conformal dimension $\D=3$, the asymptotic expansion is
\be \label{expGPPZ}
\F(x,\r)=\r^{1/2} [\f_{(0)} + \r (\f_{(2)} + \log \rho \psi_{(2)})+ ...]
\ee
The asymptotic solution can be found in the appendix.
The coefficient $\f_{(2)}$ is undetermined and only the trace
and divergence of $g_{(4)}$ are determined. The counterterms needed
in order to cancel all divergences are given by
\bea
S_{{\sm ct}}&=& \int_{\r=\e} \de^4x \sqrt{\g}
\left({3 \over 2} - {1 \over 8} R
+\half \Phi^2 \right. \\
&&\left.- \log \e
\left[{1 \over 32} (R_{ij} R^{ij} - {1 \over 3} R^2)
+{1 \over 4} (\Phi \Box_\c \Phi + {1 \over 6} R \Phi^2)\right] \right)
\nonumber
\eea
where $\Box_\c$ is the Laplacian of $\c$.
These counterterms contain (\ref{ctbkG}), as they should.
We further supplement them with the finite counterterm in (\ref{ctfinite}).
This corresponds to choosing a supersymmetric scheme.

The renormalized action is defined
\be
S_{{\sm ren}} = \lim_{\e \to 0} [S_{{\rm reg}} + S_{{\rm ct}}
+ S_{{\rm ct,fin}}]
\ee
The holographic one-point functions are equal to
\be \label{Ogppz}
\< \co \> = {1 \over \sqrt{g_{(0)}}}
{\delta S_{\sm{ren}} \over \delta \f_{(0)}} =
\lim_{\e \to 0} \left( {1 \over \e^{3/2}} {1 \over \sqrt{\g}}
{\delta S_{\sm{ren}} \over \delta \F(x,\e)} \right) =
-2 (\f_{(2)} + \psi_{(2)}) + {2 \over 9} \phi_{(0)}^3
\ee
and
\bea \label{stressGPPZ}
\<T_{ij}\>&=& \label{TGPPZ}
g_{(4)ij} +
{1 \over 8}[\Tr\, g_{(2)}^2 - (\Tr\, g_{(2)})^2]\, g_{(0) ij}
- \half (g_{(2)}^2)_{ij}  \\
&& +{1 \over 4}\, g_{(2) ij}\, \Tr\, g_{(2)}
+[\f_{(0)}(\f_{(2)} -\half \psi_{(2)}) 
-{1 \over 4} (\nabla \f_{(0)})^2]g_{(0)ij}
\nonu
&& + \half \nabla_i \f_{(0)} \nabla_j \f_{(0)}
+{3 \over 4} (T_{ij}^a + T_{ij}^\f) 
\nonumber
\eea
where $T_{ij}^a + T_{ij}^\f$ is the stress energy tensor of the 
the action equal to the integral of the trace anomaly 
$\ca=\ca_{grav} + \ca_{scal}$ (see appendix A.2). The gravitational
anomaly is given in (\ref{gravanom}) and the matter anomaly 
in (\ref{matan}) below. Although these terms are scheme
dependent and can be omitted, we prefer to work with (\ref{stressGPPZ})
keeping all the terms. 

Under the transformation (\ref{rresc}) the solution transforms 
as follows
\bea
&&\f_{(0)}'(x') = \mu \f_{(0)} (x' \mu) \nonu
&&\f_{(2)}'(x') = \mu^3 [\f_{(2)}(x'\mu) + \log \mu^2 \psi_{(2)}(x' \mu)] \\
&&g_{(0)}'(x') = g_{(0)}(x' \mu),
\qquad g_{(2)}'(x')=\mu^2 g_{(2)}(x' \mu) \nonu
&&g_{(4)}'(x') = \mu^4 [g_{(4)} + \log \mu^2 (\half T_{ij}^a 
+ \half T_{ij}^\f - \phi_{(0)} \psi_{(2)} g_{(0)})](x' \mu) \nonumber
\eea
It follows that 
\bea \label{RGT}
&&\< \co(x') \>' = \mu \left(\< \co(x' \mu) \> 
- 2 \log \mu^2 \psi_{(2)}(x' \mu) \right) \\
&&\< T_{ij}(x') \> = \mu^4 \left(
\<T_{ij}(x' \mu)\> + \half \log \mu^2 (T_{ij}^a + T_{ij}^\f)(x'\mu)\right) 
\nonumber
\eea
Exactly as in the CB case, the new terms can be obtained from 
a local finite counterterm equal to $\ca/2$, i.e. (\ref{finanom})
but with $\ca=\ca_{grav} + \ca_{scal}$ ($\ca_{grav}$ 
and $\ca_{scal}$ given in (\ref{gravanom}) and (\ref{matan}), respectively).

Evaluating these expressions on the background one obtains
\be
\<T_{ij}\>_B=0, \qquad \<\co\>_B=0
\ee
The first of these was guaranteed because we added
$S_{{\rm ct,fin}}$ to enforce
supersymmetry, but this same addition was crucial to obtain $\<O\>_B=0$.
The latter is required by the physical interpretation that the $GPPZ$
flow at $\sigma =0$ corresponds to a Lagrangian deformation of $\cn=4$
SYM without vev.  

Using the solution of the bulk field equations given in
appendix A.2 one shows that
\bea
&&\hspace{-.5cm}
\nabla^i \<T_{ij}\> = -  \<\co\> \nabla_j \f_{(0)} \label{divWGPPZ} \\
&&\hspace{-.5cm}\<T^i_i\>=- \f_{(0)} \< \co \>
+{1 \over 16} (R_{ij} R^{ij} -{1 \over 3} R^2)
-\half [(\nabla \f_{(0)})^2 - {1 \over 6} R \f_{(0)}^2]
\label{trWGPPZ}
\eea
These are the expected Ward identities.
The second term in (\ref{trWGPPZ}), is the holographic gravitational
conformal anomaly (\ref{gravanom}) \cite{HS}
and the last term,
\be \label{matan}
\ca_{scal} = - \half [(\nabla \f_{(0)})^2 - {1 \over 6} R \f_{(0)}^2]
\ee
is a conformal anomaly due to matter \cite{PeSk}.
The integrated anomaly should itself be conformal invariant,
and indeed (\ref{matan}) is equal to
the Lagrangian of a conformal scalar. The coefficients
are also the ones dictated by non-renormalization theorems.

\section{Linearized analysis around domain-walls}
\setcounter{equation}{0}

As discussed in the Introduction we must go beyond the near boundary analysis
to fix the undetermined coefficients and obtain correlation functions.
A solution of the non-linear EOM's is beyond reach, but for our
purpose of computing two-point functions it is sufficient
to consider the linearized problem.

We look for fluctuations around the solution $(A(r), \vf_B(r))$
of (\ref{1storder}). The background plus linear fluctuations is
described by
\bea
ds^2&=& e^{2A(r)}[\d_{ij} + h_{ij}(x,r)] dx^i dx^j + (1+ h_{rr}) dr^2
\nonu
\Phi&=& \vf_B(r) + \tvf(x,r)
\eea
where $h_{ij}$, $h_{rr}$ and $\tvf$ are considered infinitesimal.
This choice does not completely fix the bulk diffeomorpisms.
One can perform the one-parameter family of `gauge transformations'
\be \label{diffeo}
r= r' + \e^r(r',x'), \qquad
x^i = x'^i + \e^i(r',x')
\ee
with
\be \label{edif}
\e^i = \d^{ij} \int_r^\infty dr' e^{-2 A(r)}\pa_j \e^r
\ee
where we only display the fluctuation-independent part of $\e^i$.
These diffeos are related to those which induce the Weyl transformation
(\ref{weyl}) of the sources. The gauge choice is also 
left invariant by the linearization of the $4d$ diffeomorphisms in 
(\ref{diff4}).

We have seen in section \ref{rgsection} that the $h_{(0)ij}^L$
part of $g_{(0)ij}$ drops out from the variation of the effective
action. Equivalently the $4d$ diffeos can be fixed to
set $h_{(0)ij}^L=0$, see (\ref{diff4}).
We thus decompose the metric fluctuation as
\be
h_{ij}(x,r) = h^{T}_{ij}(x,r) + \d_{ij} {1 \over 4} h(x,r)
- \pa_i \pa_j H(x,r)
\ee
Near the boundary the fluctuations $h^{T}_{ij}(x,r),  h(x,r)$ and
$H(x,r)$ admit an
expansion similar to $g(x,\r)$. The field
theory sources are the leading $\r$-independent parts.
Analogously the scalar fluctuation $\tvf$ has the expansion given
in (\ref{expCB}) or (\ref{expGPPZ}) with source $\vf_{(0)}(x)$\footnote{
A few words about our notation:
as in (\ref{expCB}) and (\ref{expGPPZ}), $\vf_B$ and $\tvf$
have a $\r$-expansion. The corresponding coefficients will 
be denoted by $\vf_{B(2k)}$ and $\vf_{(2k)}$, $k=0,1,...$.
Then the coefficients appearing in the near-boundary analysis are given 
by $\f_{(2k)}=\vf_{B(2k)}+\vf_{(2k)}$.}.

The equation for the transverse traceless modes decouples from the
equations for $(\tvf, h, H, h_{rr})$.
The coupled graviton-scalar field equations in the axial gauge 
where $h_{rr}=0$
were derived in \cite{Notes}, and we now include $h_{rr}$. The fluctuation
equations are
\bea
&&[\pa_r^2 + 4 A' \pa_r + e^{-2A} \Box] f(x,r)=0, \ \
h^T_{ij} = h^T_{(0)ij} f(x,r) \label{tteqn} \\
&&h' = - {16 \over 3} \vf' \tvf + 4 A' h_{rr} \label{heqn1} \\
&&H'' + 4 A' H' - \half e^{-2A} h - h_{rr} e^{-2 A} =0 \\
&& 2 A' H' = \half e^{-2 A} h + {8 \over 3}{1 \over p^2} W_{\varphi}
(\tvf' - W_{\varphi\varphi} \tvf - \half W_{\varphi} h_{rr})
\eea
where $h^T_{(0)ij}$ is transverse, traceless, and independent of $r$.

The fluctuations $(\tvf, h, H, h_{rr})$ transform under the
`gauge transformations' in (\ref{diffeo}). One can form the following
gauge invariant combinations
\be \label{ginv}
R\equiv h_{rr} - 2 \pa_r \left({\tvf \over W_{\varphi}}\right), \qquad
h+{16 \over 3} {W \over W_{\varphi}} \tvf,
\qquad H'- {2 \over W_{\varphi}} e^{-2A} \tvf
\ee
In terms of these variables the equations simplify,
\bea \label{eqnsim}
&&h+{16 \over 3} {W \over W_{\varphi}} \tvf =
- {16 \over 3} {e^{2A} \over p^2}\left(
R (W W_{\varphi\varphi} - {4 \over 3} W^2 - \half W_{\varphi}^2) +\half R' W \right) \label{hfeqn}\\
&&H'- {2 \over W_{\varphi}} e^{-2A} \tvf = {1 \over p^2}
\left(2 R (W_{\varphi\varphi} - {4 \over 3} W) + R' \right)
\eea

Equation (\ref{heqn1}) takes the form
\be \label{eqh2}
\left(h+{16 \over 3} {W \over W_{\varphi}} \tvf \right)' = -{8 \over 3} W R
\ee
Differentiating (\ref{hfeqn}) leads to the
second order differential equation
\be
R''+(2W_{\varphi\varphi}-4W) R' -(4 W_{\varphi}^2 - 2 W_{\varphi}
W_{\varphi\varphi\varphi} - {32\over 9} W^2
+{8\over 3} W W_{\varphi\varphi}
+ p^2 e^{-2A}) R = 0
\ee

Equations (\ref{eqnsim})-(\ref{eqh2}) are invariant under the transformations
(\ref{diffeo}), and this is important for the success of the present
method. We will use these equations, but
henceforth consider the theory in the axial gauge,
i.e. we set $h_{rr}=0$. In \cite{Notes} the axial gauge was imposed early.
The resulting equations were equivalent to  (\ref{eqnsim})-(\ref{eqh2}) but
were processed differently. In particular because of the assumed 
conditions at the interior singularity, the boundary data for metric 
and field sources were not independent, and (at least in the GPPZ case) 
they were even non-locally related. This made it difficult to obtain
2-point functions, as one would have to disentangle the non-local 
relations of the sources from the true non-local 2-point function.

\section{Correlation functions}
\setcounter{equation}{0}

\subsection{Coulomb branch flow}

Let us now discuss the two-point functions for the trace-scalar sector.
The differential equation for $R$ becomes,
\be
R''(v)+{2 \over v} R'(v) -{p^2 \over 4 l^{2}} {1 \over v^2 (1-v)}R(v)=0
\ee
where the prime indicates derivative with respect to $v$.
The solution of this equation is
\be
R(v,p)=v^a(1-v) F(1+a,2+a,2+2a;v)
\ee
where $a=-\half (1 - \sqrt{1 + {p^2 \over l^2}})$.
Equations (\ref{eqnsim}) become
\bea \label{Couflct1}
h(v)&=&{4 \over 3} (v+2) \left( {\sqrt{6} \over (1-v)} \tvf +
{2 l^2 \over p^2} v R'\right) + {16 l^{2} \over 3 p^2} R \\
H'(v)&=& {1 \over 2 v} \left( {\sqrt{6} \over (1-v)} \tvf +
{2 l^2 \over p^2} v R'\right) + {2 l^{2} \over 3 v p^2} R \label{Couflct2}
\eea
One can solve (\ref{Couflct1}) for $\tvf_{(0)}$
(the coefficient of order $\r$ in the near-boundary expansion
of $\tvf(x,\r)$),
\be \label{relation}
\tvf_{(0)} = {1 \over 4 \sqrt{6}} h_{(0)} + \vf_{(0)}
[-\bar{Q} + {l^{2} \over p^2} (4 a^2 -{8 \over 3})]
\ee
where $\bar{Q}=\psi(1) + \psi(2)- \psi(1+a) - \psi(2+a)$ \cite{Bateman}.
It follows that
\be
\< \co \> = {N^2 \over 2 \p^2}\left(
-{2 l^2 \over \sqrt{6}} + {1 \over 2 \sqrt{6}} h_{(0)} + \vf_{(0)}
[4 \psi(1+a)-4 \psi(1) - {16 l^{2}\over 3 p^2}]\right)
\ee
Substituting this relation in (\ref{finalS}), using
$\sqrt{g_{(0)}} = 1 + \half h_{(0)} + \half p^2 H_{(0)}$,
and going to momentum space
one finds that (\ref{finalS}) can be easily integrated to
\bea \label{cbaction}
S_{{\sm ren}}[h_{(0)}, H_{(0)}, \vf_{(0)}]&=& {N^2 \over 2 \p^2}
\int d^{4}p \left(
-{1 \over 2 \sqrt{6}} \varphi_{(0)} h_{(0)}
-{p^2 \over \sqrt{6}} \varphi_{(0)} H_{(0)}  \right. \\
&& \left. +\vf_{(0)} [-{2 l^2 \over \sqrt{6}}
+\vf_{(0)} (2 \psi(1+a) -2 \psi(1)
- {8 l^{2} \over 3 p^2})] \right)
\nonumber
\eea
{}From here one immediately reads all two-point functions,
\bea
&&\<\co(p) \co(-p)\> = {N^2 \over 2 \p^2}
[4 \psi(1) - 4 \psi(1+a) + {16 l^2 \over 3 p^2}]
\label{OOCB} \\
&&\<T^i_i(p) \co(-p)\> = -{4 l^2 \over \sqrt{6}} {N^2 \over 2 \p^2}
= 2 \<\co\>_B \label{TOCB}\\
&&p^i p^j \<T_{ij}(p) \co(-p)\> = 
-{2 l^2 \over \sqrt{6}} {N^2 \over 2 \p^2} p^2
=  \< \co \>_B p^2 \label{dTCB}
\eea
These correlation functions, as well as the one in (\ref{TTCB}),
correctly approach the corresponding 
$AdS$ correlators, including normalizations,
in the limit of vanishing vev, i.e. $l \to 0$.
Notice that the contact term in
(\ref{TOCB}) and (\ref{dTCB}) are exactly the ones dictated
by the Ward identities (\ref{trWCB}) and (\ref{divWCB}).

It remains to discuss the tranverse traceless sector.
To determine the two-point function we need to obtain $g_{(4)ij}$
as a function of $h^T_{(0)ij}$.
The solution of the fluctuation equation (\ref{tteqn}) is given by
\cite{fgpw2}
\be
f(v,p)=v^a F(a,a,2+2a;v)
\ee
Expanding around $v=1$ and converting to the $\r$ variable
one obtains
\be
h_{(4)ij}^T =  h^T_{(0)ij} \left({p^4 \over 32} [\psi(1) + \psi(3)]
+{p^2 l^{2}\over 24} - {p^4 \over 16} \psi(a+1) \right)
\ee
The first term on the right hand side yields only a contact
term and can be omitted when computing correlators at
non-coincident points. From the linearization of (\ref{tij1}) we obtain
the correlation function 
\be \label{TTCB}
\<T_{ij}(p) T_{kl}(-p)\>= \P^{TT}_{ijkl} {N^2\over 2 \pi^2}
\left({p^2 l^2 \over 12 } - {p^4 \over 8} \psi(a+1) \right)
\ee
Both (\ref{OOCB}) and (\ref{TTCB}) contain 0-mass poles. In(\ref{TTCB}) 
the projection operator gives the pole term $p_ip_jp_kp_l/p^2$.
These poles have physical sign. Their contribution dominates the
long-distance behavior of the Euclidean correlation functions and
obeys reflection positivity. The poles are a manifestation of the
expected dilaton, the Goldstone boson of spontaneously broken
conformal symmetry. Pole residues are proportional to the vev 
$\<\co\>_B$.  
In addition to the massless poles, there is a branch cut along the positive
real axis indicating a continuous spectrum above the mass scale
\be \label{mW}
m_W={l \over \a'}
\ee
in exact agreement with gauge theory expectations:
$l/\a'$ is the average mass of the $W$-bosons \cite{fgpw2},
and the vev that breaks the symmetry is proportional
to $l^2$. The connected correlator (see Footnote \ref{QFT})  
\be
\<T_{ij}(p) \co (-p)\> = {4 l^2 \over 3\sqrt6} {N^2 \over 2 \p^2} \pi_{ij}
=-{2 \over 3} \< \co \>_B \p_{ij}
\ee
has the dilaton pole but is otherwise purely local. Furthemore,
\be
\<T^i_i(p)T^j_j(-p)\> =0
\ee
as can be seen from (\ref{cbaction}) or from
the linearization of (\ref{trWCB}). These new results from supergravity 
are fully consistent with spontaneously broken conformal symmetry in
the dual field theory!

\subsection{$GPPZ$ flow}

The equation for $R$ becomes
\be
R'' + {1 \over u (1-u)} \left[ R' +  \left({2u-1 \over u (1-u)} - {p^2 \over
4}\right)  R \right]=0
\ee
where the prime indicates differentiation with respect to $u$.
The solution of this equation is
\be
R(u,p)=u (1-u) F({3 \over 2} +\half b, {3 \over 2} -\half b, 3;u)
\ee
where $b=\sqrt{1 - p^2}$.

Using the definition of $R$ in (\ref{ginv}) we find
\be \label{goofball}
\vf_{(2)} + \psi_{(2)} = \vf_{(0)}
- p^2 ({1 \over 4} \vf_{(0)} + {\sqrt{3} \over 32} h_{(0)}) \bar{J}
\ee
where we used
$\psi_{(2)}=p^2 ({1 \over 4} \vf_{(0)} + {\sqrt{3} \over 32} h_{(0)})$
from (\ref{gppzdata}) and $\bar{J}=2 \psi(1) - \psi({3 \over 2} +\half b)
-\psi({3 \over 2} - \half b)$ from \cite{Bateman}. The linearization
of (\ref{Ogppz}) then gives
\be
\<\co\>=
-2 (\vf_{(2)} + \psi_{(2)}) + {2 \over 3} \vf_B^2 \vf_{(0)}=
p^2({1 \over 2} \vf_{(0)} + {\sqrt{3} \over 16} h_{(0)}) \bar{J}
\ee

Inserting $\<\co\>$ in (\ref{finalS}) we obtain the action
\be
S_{{\sm ren}}[h_{(0)},H_{(0)},\vf_{(0)}]= {N^2 \over 2 \p^2}
\int d^4p \left( 
{1 \over 4} (\vf_{(0)} + {\sqrt{3} \over 8} h_{(0)})^2 p^2 \bar{J} 
+{3 \over 256} p^2 h_{(0)}^2
\right)
\ee
where we have also included the overall factor $N^2/2 \p^2$.
This action gives all correlators of $T_i^i$ and $\co$. The 
last term is due to the anomaly.
It follows that there is the operator relation,
\be
T^i_i = \beta_{\co} \co
\ee
where $\beta_{\co}=-\varphi_{(0)B}=-\sqrt{3}$. We will discuss this below.

The correlation functions are equal to
\bea
&&\<\co(p) \co(-p)\> = -{1 \over 2} {N^2 \over 2 \p^2} p^2 \bar{J}
\label{OOGPPZ} \\
&&\<T^i_i(p) \co(-p)\> = {\sqrt{3} \over 2} p^2 {N^2 \over 2 \p^2} \bar{J} \\
&&\<T^i_i(p) T^i_i(-p)\> = -{3 \over 2} {N^2 \over 2 \p^2} p^2 (\bar{J} +1)
\label{trgppz} 
\eea
The correlator $\< \co (p)  \co (-p)\>$ 
agrees with \cite{aft}. The shift from $\bar{J}$
to $\bar{J}+1$ in (\ref{trgppz}) is due to the linearization of the
trace anomaly term in (\ref{trWGPPZ}). There is then a cancellation, 
and the correlator vanishes at the rate $p^4$ at low momentum. This 
is relevant to the issue of 0-mass poles discussed below. 

We now turn to the transverse traceless correlator.
The solution to the fluctuation equation is \cite{agpz,Notes}
\be
h_{ij}^T(u,p) \propto (1-u)^2 F(2+{ip \over 2}, 2-{ip \over 2},2;u) 
 h_{(0)ij}^T(p)
\ee
which has the asymptotic expansion \cite{Bateman}
\be \label{ttexpan}
h_{ij}^T = [1 -\r{p^2 \over 4}  + \r^2 {1 \over 32} p^2 (4 + p^2) 
(\bar{K}-\log \r) + \cdots]  h_{(0)ij}^T
\ee
where $\bar{K}=\psi(1) + \psi(3) -\psi(2+{ip \over 2})-
\psi(2-{ip \over 2})$.
After careful linearization of (\ref{TGPPZ}) one finds the following 
contribution to the transverse traceless correlator
\be \label{lTij}
\<T_{ij}\> = g_{(4)ij} -{3 \over 16} p^2 h_{(0)ij}^T= 
h_{(4)ij}-h_{(2)ij}-{3 \over 16} p^2 h_{(0)ij}^T
\ee
where $h_{(2)}$ and $h_{(4)}$ are the order $\r$ and $\r^2$ terms in
the expansion (\ref{ttexpan}). The first term in (\ref{lTij}) 
is the linearization of $g_{(4)ij}$ in (\ref{stressGPPZ}); 
the second is a linear correction to $g_{(4)ij}$ arising
because the background scale factor in (\ref{solrho}) has an order 
$\rho$ contribution; finally the last term comes from the linearization 
of the term $3 T_{ij}^{\f} /4$ in (\ref{stressGPPZ}).
The desired correlation function is then
\be \label{TTGPPZ}
\<T_{ij}(p) T_{kl}(-p)\>_{TT}= \P^{TT}_{ijkl} {N^2\over 2 \pi^2}
[{1 \over 16} p^2 (p^2 + {4}) \bar{K} + {p^2 \over 8}]
\ee
To this one must add the trace part obtained from (\ref{trgppz})
\be \label{longgppz}
\<T_{ij}(p) T_{kl}(-p)\>_{trace}= -\p_{ij}\p_{kl} {N^2\over 2 \pi^2}
p^2 {(\bar{J}+1) \over 6}
\ee

We now discuss some of the physical aspects of our results. First we
note that the correlators themselves are not completely new, 
\cite{agpz,aft,pstarinets}. 
Rather, it is
the coherent derivation via the holographic renormalization procedure
which is to be emphasized. 

Physically one should expect that correlators become insensitive to the
shape of the domain wall as $p^2 \rightarrow \infty$ and have the same
limiting form $p^{2\Delta-4} \log p$ as those in a pure $AdS_5$ background.
This property is satisfied by all $CB$ and $GPPZ$ correlators in this
paper. With due care for conventions, this limit may be used to normalize
all results.

The $\b_{\co}$ function which appears is a trivial
constant. In fact since $\varphi_{(0)B} = \sqrt3$, it agrees with the
classical value\footnote{See also Sec. 12.5 of \cite{peskin}.} in 
(\ref{trace}).
This is exactly what we expect here, since the $GPPZ$ flow is dual to
the deformation of $\cn =4$ SYM superpotential by the $\cn =1$ 
supersymmetric mass terms
$W={ m \S_i \Tr (\Phi_i^2) }$ with $m=\varphi_{(0)B}$. The $\cn=1$ 
non-renormalization theorem implies that $m$ is renormalized only through
anomalous dimension of the operator it multiplies.
However, in our case
$W$ is a protected operator in the undeformed theory, and there is 
no anomalous dimension\footnote{
Contrary to naive expectation, the supersymmetric mass term does not involve
the lowest scalar operator ${\cal K} = \S_i \Tr (\phi_i^2)$ 
in the $\cn = 4$ Konishi multiplet \cite{afz}, with non-zero
anomalous dimension at weak coupling \cite{afz,andim} that
grows as $(g_sN)^{1/4}$ at strong coupling. Indeed  
the Konishi scalar is the simplest among the
operators dual to string excitations \cite{gkp,afz} not
captured by the supergravity limit.}. The $\b_{\co}$ function
is indeed just classical. Notice that our  $\b_{\co}$ function is 
equal to the leading term
in the $\r$ expansion of the ``holographic'' beta function
of \cite{Behr,dBVV}.

In general our correlation functions have the discrete spectrum of poles
noted in earlier discussions of the $GPPZ$ flow. However there is a
more delicate question of a 0-mass pole. One might have naively
expected that terms in the invariant amplitudes 
$A(p^2)$ and $B(p^2)$ that vanish in the limit $p^2 \to 0$
are just contact terms. However, because the correlators
involve projection operators, terms that vanish at the
rate $p^2$ yield zero-mass poles instead.
The presence of such a zero-mass pole would conflict with the
interpretation that the $GPPZ$ flow is dual at long distance to
$\cn =1$ SYM which is a confining theory with mass gap.
We find that both $A(p^2)$ and $B(p^2)$ vanish as $p^4$
at low momentum, and thus the correlators do not contain 
zero-mass poles. 

One may ask whether our results 
depend on the specific radial coordinate used, specifically the
scaling $\r \rightarrow \m^2 \r$ associated with an $RG$ transformation
in Section 3. It was shown there that the rescaling 
introduces
new terms in the stress-energy tensor which
can be derived from the anomaly action which is local and thus 
produces only contact terms in correlation functions such as the complete
$\<T_{ij}T_{kl}\>$. Since our method computes the $TT$ and trace parts
of $\<T_{ij}T_{kl}\>$ separately one can see from hypergeometric
series such as (\ref{ttexpan}) that terms of order $p^2 \log \m^2$ appear.
These potentially give 0-mass poles. However,
one can check explicitly that the 0-mass poles cancel between the $TT$
and trace parts of the correlator, leaving only the net $p^4 \log \m^2$
contact term from the anomaly.
 
\section{Conclusion}
In this paper we have developed a coherent approach to correlation
functions of the stress tensor in holographic $RG$ flows. The
implementation of holographic renormalization is somewhat tedious,
but valid for all solutions of a given bulk action. The procedure
gives relatively simple finite expressions for source dependent vevs 
$\<T_{ij}\>$ and $\<\co\>$ from which
correlation functions can easily be obtained.
Ward identities with correct anomalies are satisfied. 
The expected physics of flows describing
both operator and Coulomb deformations of $\cn=4$  SYM is reflected
in the results. There is more to be done. The correlation
functions presumably satisfy Callan-Symanzik equations, see \cite{Erd}
for a recent discussion. The treatment
should be extended to other important field theory operators such as vector  
currents. We hope to discuss these questions and present more of the
technical details of the procedure in \cite{later}.

\section*{Acknowledgement} 
The authors acknowledge recent e-mail from W. M\"uck indicating that he
also has results on the correlation functions studied above.
Various physics centers provided stimulation
and support during the investigation which led to this paper. These
include the  Centre Emile Borel (for DZF), the Aspen Center for Physics,
UCLA Department of Physics, and CIT-USC Center for Theoretical Physics 
(for DZF and KS). The research of KS is supported in part by the
National Science Foundation grant PHY-9802484. 
The research of DZF is supported in part by the
National Science Foundation grant PHY-9722072. The research of MB is
supported in part by the EEC contract HPRN-CT-2000-00122, a PPARC grant and
the INTAS project 991590. 
The research of MB and DZF is partially supported by the INFN-MIT
``Bruno Rossi'' exchange.


\appendix

\section{Asymptotic solutions}
\setcounter{equation}{0}

\subsection{Coulomb branch flow}

Scalar field:
\bea \label{feqs}
&&\f_{(2)}=-{1 \over 4} \left( \Box_0 \f_{(0)}
+ {2 \over 3} \f_{(0)} R[g_{(0)}] \right)
- {4 \over \sqrt{6}} (\f_{(0)}^2 - \half \f_{(0)} \tf_{(0)}) \nonu
&&\tf_{(2)}=-{1 \over 4} \left( \Box_0 \tf_{(0)}
+ {1 \over 3} R[g_{(0)}] (\tf_{(0)} + \f_{(0)}) + 8 (\f_{(2)} + \psi_{(2)})
\right)+ {1 \over \sqrt{6}} \tf_{(0)}^2 \nonu
&&\psi_{(2)}= {1 \over \sqrt{6}} \f_{(0)}^2
\eea

Metric:
\bea \label{cbdata}
&&g_{(2)}{}_{ij} =  \frac{1}{2} \left( R_{ij} - \frac{1}{6}
R\, g_{(0)ij} \right), \nonu
&&h_{1(4)ij}=h_{(4)ij} - {2 \over 3} \f_{(0)} \tf_{(0)} g_{(0)ij}, \qquad
h_{2(4)ij}=-{1 \over 3} \f_{(0)}^2 g_{(0)ij} \nonu
&& \Tr g_{(4)} = {1 \over 4} \Tr g_{(2)}^2
-{2 \over 3} (\f_{(0)}^2 + 2 \tf_{(0)}^2), \nonu
&& \nabla^j g_{(4) ij} = \nabla^j
\left(- {1 \over 8}[\Tr\, g_{(2)}^2 - (\Tr\, g_{(2)})^2]\, g_{(0) ij}
+ \half (g_{(2)}^2)_{ij}
\right. \nonu
&&\left.
\qquad - {1 \over 4}\, g_{(2) ij}\, \Tr\, g_{(2)}
-{1 \over 3} (2 \f_{(0)}^2 + \tf_{(0)}^2-3\f_{(0)} \tf_{(0)}) g_{(0)ij}\right)
\nonu
&& \qquad
-  2 \tf_{(0)} \nabla_i \f_{(0)}
\label{div}
\eea
The tensor $h_{(4)}$ is equal to \cite{KSS}
\bea 
h_{(4)}&=&{1 \over 8} R_{ikjl} R^{kl} + {1 \over 48} \nabla_i \nabla_j R
-{1 \over 16} \nabla^2 R_{ij} -{1 \over 24} R R_{ij} \nonu
&&+ ({1 \over 96} \nabla^2 R + {1 \over 96} R^2 -{1 \over 32} R_{kl}R^{kl}) 
\nonu
&=& \half T_{ij}^a \label{h4}
\eea
where $T_{ij}^a$ is the stress energy tensor derived from the action 
\be
S_a=\int \de^4x \sqrt{g_{(0)}} \ca_{grav} \ .
\ee
$\ca_{grav}$ is the gravitaional trace anomaly given in (\ref{gravanom}).

\subsection{$GPPZ$ flow}
Scalar field:
\be \label{scdata}
\psi_{(2)} = - {1 \over 4} (\Box_0 \f_{(0)} + {1 \over 6} R \phi_{(0)})
\ee

Metric:
\bea \label{gppzdata}
&&g_{(2)}{}_{ij} =  \frac{1}{2} \left( R_{ij} - \frac{1}{6}
R\, g_{(0)ij} \right) - {1 \over 3} \f_{(0)}^{2} g_{(0)ij} \nonu
&&h_{1(4)ij}=h_{(4)ij}+{1 \over 12} R_{ij} \f_{(0)}^2
-{1 \over 3} \nabla_i \f_{(0)}\nabla_j \f_{(0)}
+{1 \over 12} (\nabla \f_{(0)})^{2} g_{(0)ij} \nonu
&&\qquad \qquad
+{1 \over 6} \f_{(0)} \nabla_i \nabla_j \f_{(0)}
+ {1 \over 12} \f_{(0)} \Box_0 \f_{(0)} g_{(0)ij} \nonu
&&\qquad \qquad = h_{(4)ij} + {1\over 2} T^{\f}_{ij}
+ {1 \over 4}
g_{(0)ij} (\f_{(0)} \Box_0 \f_{(0)} + {1 \over 6} R \phi_{(0)}^2) \nonu
&&h_{2(4)ij}=0, \nonu
&&\Tr g_{(4)} = -2 \f_{(0)} \f_{(2)} - {1 \over 4}  \f_{(0)} \Box_0 \f_{(0)}
-{5 \over 72} R \f_{(0)}^2 \nonu
&&\qquad \qquad
+ {1 \over 16}(R_{ij} R^{ij} -{2 \over 9} R^2) + {2 \over 9} \f_{(0)}^4 \nonu
&&\nabla^j g_{(4)ij}= \nabla^j
\left(- {1 \over 8}[\Tr\, g_{(2)}^2 - (\Tr\, g_{(2)})^2]\, g_{(0) ij}
+ \half (g_{(2)}^2)_{ij}
\right. \nonu
&&\qquad \qquad
- {1 \over 4}\, g_{(2) ij}\, \Tr\, g_{(2)}
-{3 \over 2} h_{1(4)ij} - \half \nabla_i \f_{(0)} \nabla_j \f_{(0)}
 \nonu
&&\left.\qquad \qquad
+ g_{(0) ij}\left[{1 \over 4} (\nabla \f_{(0)})^2 - \f_{(0)} (\f_{(2)}
+ \psi_{(2)})\right]  \right) \nonu
&&\qquad \qquad
- [-2 (\f_{(2)} + \psi_{(2)}) + {2 \over 9} \f_{(0)}^3] \nabla_i \f_{(0)}
\eea
where $T^{\f}_{ij}$ is the stress energy tensor derived from the
action
\be
S_\f = \int \de^4x \sqrt{g_{(0)}} \ca_{scal}
\ .
\ee
$\ca_{scal}$ is the matter conformal given in (\ref{matan}). 
$h_{(4)}$ is given 
by the same formula (\ref{h4}) as in the CB case. 


\end{document}